\documentstyle[12pt]{article}                                 
\begin{document}

\newcommand {\bea}{\begin{eqnarray}}
\newcommand {\eea}{\end{eqnarray}}
\newcommand {\be}{\begin{equation}}
\newcommand {\ee}{\end{equation}}

\title{
\begin{flushright}
\begin{small}
hep-th/9705192 \\
UPR/752-T \\
May 1997 \\
\end{small}
\end{flushright}
\vspace{1.cm}
General Rotating Black Holes in String Theory: Greybody Factors
and Event Horizons}

\author{Mirjam Cveti\v{c} and Finn Larsen\\
%\thanks{Research supported in part by 
%Danish National Science Foundation.}\\
\small Department of Physics and Astronomy\\
\small University of Pennsylvania\\
\small Philadelphia, PA 19104 \\
\small e-mail: cvetic,larsen@cvetic.hep.upenn.edu
}
\date{ }
\maketitle

\begin{abstract}
We derive the wave equation for a minimally coupled scalar field in 
the background of a general rotating five-dimensional black hole. 
It is written in a form that involves two types of thermodynamic 
variables, defined at the inner and outer event horizon, respectively. 
We model the microscopic structure as an effective string theory, 
with the thermodynamic properties of the left and right moving excitations 
related to those of the horizons. Previously known solutions to the 
wave equation are generalized to the rotating case, and their regime 
of validity is sharpened. We calculate the greybody factors and interpret 
the resulting Hawking emission spectrum microscopically in several 
limits. We find a U-duality invariant expression for the effective 
string length that does not assume a hierarchy between the charges.
It accounts for the universal low-energy absorption cross-section
in the general non-extremal case.

\end{abstract}                                
\newpage
\section{Introduction}
\label{sec:intro}

Hawking's seminal calculation of the black hole temperature allows
for a surprising window to quantum gravity: it immediately yields the
size of the underlying space of quantum states in quantitative 
detail~\cite{hawking75}. The result relies only on a particular detail 
of the black hole geometry, namely its limiting form close to the 
{\it outer} event horizon. We will argue that other geometric properties 
give similarly direct evidence on the microscopic structure of black 
holes. Specifically, we find an important role for the geometry in the 
vicinity of the {\it inner} event horizon, as well.

The discussion and the examples aim at the description of black 
holes as quantum states in string theory (for review 
see~\cite{juanthesis,horowitz97}). It is a characteristic property 
of string models that the entropy is the sum of contributions from left 
and right moving excitations of the string; and the thermodynamic 
variables accordingly appear in duplicate versions. 
The black hole geometry exhibit an analogous structure: standard 
thermodynamic variables, defined at the outer event horizon, are  
mirrored by an independent set of thermodynamic variables, defined at 
the inner event horizon. We find that the left and right moving 
thermodynamics of the string theory corresponds to the sum 
and the difference of the outer and the inner horizon thermodynamics.
This relation can be established by direct inspection for large classes 
of extremal and near-extremal black holes. Indeed, it is valid in all 
the cases where the correspondence between black holes and string 
theory has been demonstrated. Ultimately we would like to find a 
microscopic description of {\it all} black holes within
string theory; and our geometrical observations may be sufficiently 
robust to serve as guidance towards this goal (other attempts 
include~\cite{strom96d,susskind96,polch96}).

In the following we give an outline the paper and summarize the 
results in more detail.

We begin with an important motivating fact that concerns the entropy of 
general rotating black holes in five dimensions~\cite{cy96b}:
\be
S= 2\pi [\sqrt{{1\over 4}\mu^3 
(\prod_{i=1}^3 \cosh\delta_i+\prod_i \sinh\delta_i)^2-J^2_L}
+\sqrt{{1\over 4}\mu^3 
(\prod_{i=1}^3 \cosh\delta_i-\prod_i \sinh\delta_i)^2-J^2_R}]
\label{eq:bhentropy}
\ee
(As we explain in sec.~\ref{sec:thermo} the non-extremality 
parameter $\mu$ and the boosts $\delta_i$ parametrize 
the mass and the charges; and $J_{L,R}$ are angular momenta.) The 
form of the entropy may be interpreted as an indication that it 
derives from two independent microscopic contributions; and each of 
these may be attributed to a gas of 
strings~\cite{cy96b,cveticreview,fl97}. 
We will consider the general case of rotating black holes because 
the crucial division into two terms becomes ambiguous in the 
limit of vanishing angular momenta.
We develop the thermodynamics of this interpretation in detail,
in sec.~\ref{sec:thermo}. An important feature is that we find 
two independent temperatures $T_R$ and $T_L$, one for each gas. These 
two temperatures play central roles in subsequent sections.

In sec.~\ref{sec:waveeq} we present our main technical result:
we write the exact wave equation for a minimally coupled scalar in 
the most general black hole background in five dimensions
(eq.~\ref{eq:geneq}). The wave equation has a surprisingly symmetric
structure, given the 
generality of the setting. A characteristic feature 
is that the outer and inner event horizons appear in a symmetric 
fashion. The modes in the vicinity 
of the outer horizon give rise to the Hawking radiation, with 
characteristic temperature $T_H^{-1}={1\over 2}(T_R^{-1}+T_L^{-1})$. 
Analogously, from the modes in the vicinity of the inner horizon we infer
a ``temperature'' given by  $T_{-}^{-1}={1\over 2}(T_R^{-1}-T_L^{-1})$.
The temperatures $T_R$ and $T_L$ that appear in these formulae agree 
precisely with those that follow from thermodynamics.
Similar results are derived for the other thermodynamic variables, 
{\it i.e.} rotational velocities and $U(1)$-potentials.

The wave equation has an exact symmetry that interchanges the inner 
and outer event horizons. In sec.\ref{sec:symmetry} we 
identify this discrete symmetry with the T-duality of an underlying 
string theory. Moreover, we exhibit an approximate 
$SL(2,R)_R\times SL(2,R)_L$ symmetry group that is realized directly on the 
macroscopic fields. From the quantum numbers of the symmetry group we 
recover the temperatures $T_R$ and $T_L$. Although the precise 
interpretation of these facts remains unclear it is interesting that 
they point rather specifically towards a string theory description.

In sec.\ref{sec:solutions} we find solutions to the radial wave 
equation in two regions, solving first in the asymptotic region and 
then in the near horizon region. We also discuss the angular equation. 
These results generalize previously
known results to the case of rotating black holes. We discuss the ranges 
of charge, angular momenta, and mass for which these solutions can 
be combined to approximate wave functions covering the entire 
spacetime; and so the black hole absorption cross-sections can be 
calculated explicitly. The results presented in sec.~\ref{sec:abs} 
include:
\begin{itemize}
\item
The low energy S-wave absorption cross-section is:
\be
\sigma_{\rm abs}(\omega\rightarrow 0 )=A~,
\ee
where $A$ is the area of the black hole. Our result shows
that this holds for {\it all} five dimensional black holes in 
toroidally compactified string theory.
\item
For a range of parameters (that we specify) black 
holes exhibit the S-wave absorption spectrum:
\be 
\sigma_{\rm abs}^{(0)}(\omega)= A 
{{\omega\over 2T_L}{\omega\over 2T_R}\over{\omega\over T_H}}
{(e^{\omega/T_H}-1)
\over (e^{\omega\over 2T_L}-1)(e^{\omega\over 2T_R}-1)}~.
\ee
This spectrum is a precise indication that the Hawking emission process 
of the black hole can be described in an effective string
theory as a simple two-body 
process~\cite{callan96a,greybody,mathur}. In this dynamical model the 
distribution functions of the colliding quanta are thermal with the 
temperatures $T_R$ and $T_L$. The freedom afforded by the angular momenta 
allows a demonstration of this characteristic behavior in several 
regions of parameter space that were previously out of reach. For example 
there is a parameter range with {\it no} hierarchy in the relative 
magnitudes of the charges. 
\item
For a larger range of black hole parameters, and for higher partial 
waves, an explicit solution can still be found~\cite{mathur97,strominger97a}. 
In this case the absorption cross-section has a more complicated 
form and the Hawking radiation cannot be interpreted as a two-body 
process. However, it is suggestive that the emission spectrum still 
takes a factorized form where 
each factor depends on $T_R$ and $T_L$, respectively.
\end{itemize}

We complete the paper, in sec.~\ref{sec:discussion}, with a 
discussion of the microscopic description of the dynamics. 
It is shown that, for the most general black holes, the two-body 
emission processes can be modelled by a simple value of the effective 
string length. However, we also stress that, for generic non-extremal 
black holes, the typical Hawking process can not be described in this 
simple fashion.

\section{Thermodynamics of Rotating Black Holes}
\label{sec:thermo}
We are interested in a class of black holes in five dimensions that are 
parametrized by their mass $M$, 2 angular momenta $J_{R,L}$, and 3 
independent $U(1)$ charges $Q_i$~\cite{cy96a}. 
These are the most general solutions to the low energy effective
action of the heterotic and type II string theories, toroidally
compactified to five dimensions\footnote{We write formulae in their
generating form; so they are only the most general up to duality.
However, they can be written in a manifestly duality invariant 
way~\cite{gaida}.}~\cite{hull96}. The explicit expressions for 
these black holes are involved and given in detail in~\cite{cy96a}. 
For the sake of completeness
we present their spacetime metric in the appendix~\ref{app:bhsolution}. 
In this section we discuss their thermodynamical properties.

The mass and the charges of the black holes are conveniently
given in the parametric 
form\footnote{The notation here is $\mu=2m$ where $m$ is the notation 
in~\cite{cy96a}; or $\mu=r^2_0$ where $r_0$ is the
notation of~\cite{strom96d}. We choose duality invariant units where the five 
dimensional gravitational coupling constant is $G_5={\pi\over 4}$. 
In string conventions this amounts to 
$(\alpha^\prime)^4 g^2/(R_1 R_2 R_3 R_4 R_5)=1$.}:
\bea
M &=& {1\over 2}\mu\sum_{i=1}^3 \cosh 2\delta_i~, \\
Q_i &=& {1\over 2}\mu \sinh 2\delta_i~~~;~~i=1,2,3~.
\eea
The BPS-saturated limit corresponds to $\mu\rightarrow 0$ and 
$\delta_i\rightarrow\infty$ with $Q_i$ kept fixed; so
$\mu$ is a measure of the deviation from the BPS case. 
The parameters $\delta_i$ are referred to as boosts because of their 
role in the solution generating technique employed to find the 
charged black holes.

In 5 dimensions the rotation group is $SO(4)\simeq SU(2)_R\times SU(2)_L$.
Therefore black holes are characterized by two independent projections
of the angular momentum vector. These parameters are the two angular 
momenta that will be denoted $J_{R}$ and $J_{L}$. Normalizations 
have been chosen such that $J_{R,L}$ are pure numbers 
(in units where $\hbar=1$ ) that are quantized 
in the microscopic theory\footnote{The quantization condition 
is that $J_{R,L}={1\over 2}(J_\phi\pm J_\psi)$ where $J_\phi$ and $J_\psi$
are quantized as integers.}. It is sometimes convenient to parametrize 
the angular momenta of the general black hole in terms of 
the $l_{1,2}$ defined through:
\be
J_{R,L}= {1\over 2}\mu (l_1\pm l_2)(\prod_i \cosh\delta_i \mp
\prod_i \sinh\delta_i)~.
\label{eq:l12def}
\ee
The $l_{1,2}$ are the angular momenta of the Kerr black hole used as a 
starting point of the generating technique.  We will give the formulae 
in terms of $l_{1,2}$ along with those using $J_{R,L}$; because 
both forms will be needed.

\subsection{Entropy}
The black hole entropy (eq.~\ref{eq:bhentropy}) was derived 
in~\cite{cy96b}. As noted already in the introduction the entropy 
clearly divides into two terms. We make this manifest by writing 
$S=S_L+S_R$ where:
\bea
S_L&=& 2\pi \sqrt{{1\over 4}\mu^3 
(\prod_i \cosh\delta_i+\prod_i \sinh\delta_i)^2-J^2_L} \nonumber \\
&=&\pi\mu (\prod_i \cosh\delta_i+\prod_i \sinh\delta_i)
\sqrt{\mu - (l_1-l_2)^2}~. \\
S_R&=& 2\pi 
\sqrt{{1\over 4}\mu^3 
(\prod_i \cosh\delta_i-\prod_i \sinh\delta_i)^2-J^2_R} \nonumber \\
&=&\pi\mu (\prod_i \cosh\delta_i-\prod_i \sinh\delta_i)
\sqrt{\mu - (l_1+l_2)^2}~. 
\eea
By now there are many hints from string theory that collective 
excitations of solitonic objects can be described by effective low 
energy theories that are themselves string theories. The structure 
of the entropy as a sum of two terms may be an indication that {\it all} 
black holes can be described in this way; and that the two terms 
in the entropy are the contributions from left (L) and right (R) moving 
modes, respectively. If true, it must be that the interactions between 
the two kinds of modes 
can be treated as weak. Motivated by the BPS-saturated case we assume that 
the relevant effective theory is a noncritical string theory with 
$c=6$~\cite{sv1,dvv1,rotation1}; and identify the levels of the 
effective string as:
\bea
N_L&=& {1\over 4}\mu^3 
(\prod_i \cosh\delta_i+\prod_i \sinh\delta_i)^2-J^2_L~, \\
N_R&=& {1\over 4}\mu^3 
(\prod_i \cosh\delta_i-\prod_i \sinh\delta_i)^2-J^2_R~,
\eea
so that for large levels: 
\be
S=S_L+S_R=2\pi (\sqrt{N_L}+\sqrt{N_R})~.
\ee
If these relations could be derived from first principles we would have 
a microscopic interpretation of the entropy in the general non-extremal 
case. Some evidence in this direction was presented in~\cite{fl97}.

Black holes in four dimensions have entropies of a very similar
form~\cite{cy96b}: the index $i=1,2,3\rightarrow i=1,2,3,4$, 
the parameter $\mu^3\rightarrow\mu^4$, and the angular momentum $J_L=0$. 
Therefore the thermodynamics, and indeed most results presented
in this paper, immediately carry over to four dimensions.
Note however that there is only one angular momentum in four 
dimensions; so the symmetry between the two entropies 
$S_{R,L}$ is a special property of the five dimensional case that hints at 
a particularly symmetric underlying structure. We will discuss
rotating black holes in four dimensions in a separate paper~\cite{cl97b}.

\subsection{Thermodynamics}
Our assumption that the entropy is a sum of two independent contributions
has consequences that can be derived from general principles. Consider
the first law of thermodynamics:
\be
dM=T_H dS+\Omega^{R}dJ_R+
\Omega^{L}dJ_L+\sum_i\Phi^i dQ_i~.
\ee
We write the inverse Hawking temperature as:
\be 
\beta_H\equiv{1\over 2}(\beta_L+\beta_R)~,
\ee
and use $S=S_L+S_R$. Then we find:
\bea
[-{1\over 2}\beta_R dM+dS_R+\beta_H\Omega^{R}dJ_R
+\beta_H\sum_i\Phi^i_R dQ_i]+
\nonumber 
\\+
[-{1\over 2}\beta_L dM+dS_L+\beta_H\Omega^{L}dJ_L+
\beta_H\sum_i\Phi^i_L dQ_i]=0~.
\eea
The two independent inverse temperatures follow directly from this relation:
\bea
\beta_L&=& 
{\pi\mu^2 (\prod_i \cosh^2\delta_i-\prod_i \sinh^2\delta_i)
\over
\sqrt{{1\over 4}\mu^3 (\prod_i \cosh\delta_i+\prod_i \sinh\delta_i)^2-J^2_L}}
={2\pi\mu (\prod_i \cosh\delta_i-\prod_i \sinh\delta_i)\over
\sqrt{\mu - (l_1-l_2)^2}}~,
\label{eq:betal}
\\
\beta_R &=& 
{\pi\mu^2 (\prod_i \cosh^2\delta_i-\prod_i \sinh^2\delta_i)
\over\sqrt{{1\over 4}\mu^3 
(\prod_i \cosh\delta_i-\prod_i \sinh\delta_i)^2-J^2_R}}
={2\pi\mu (\prod_i \cosh\delta_i+\prod_i \sinh\delta_i)\over
\sqrt{\mu - (l_1+l_2)^2}}~.
\label{eq:betar}
\eea
In the string theory interpretation these are the physical temperatures 
of the left and right moving modes. For this to make sense we must assume 
that the modes are interacting in such a way that the thermal equilibrium 
is maintained in each of the two gasses independently; and so that 
the coupling between the two sectors are much weaker that the ones that 
act within each sector. Although this is perhaps surprising from the string 
theory point of view it may be reasonable when considering the 
nature of black holes: colliding left and right modes give rise to 
Hawking radiation, and we know that large black holes are exceedingly 
stable objects.

The angular velocities also follow from the first law
of thermodynamics:
\bea
\beta_H\Omega^L &=& {2\pi J_L\over \sqrt{{1\over 4}
\mu^3 (\prod_i \cosh\delta_i+\prod_i \sinh\delta_i)^2-J^2_L}}
={2\pi (l_1-l_2)\over\sqrt{\mu-(l_1-l_2)^2}}~,
\label{eq:omegal}
\\
\beta_H\Omega^R &=& {2\pi J_R\over \sqrt{{1\over 4}
\mu^3 (\prod_i \cosh\delta_i-\prod_i \sinh\delta_i)^2-J^2_R}}
={2\pi (l_1+l_2)\over\sqrt{\mu-(l_1+l_2)^2}}~.
\label{eq:omegar}
\eea
As before these potentials can be attributed to their respective
independent sets of modes. Note however that the inverse temperature 
$\beta_H$ is the sum of left and right contributions; 
so the rotational velocities $\Omega^{L,R}$ can not be unambiguously 
associated with a specific sector. It is only
the combinations $\beta_H\Omega^{L,R}$ that can be interpreted in this
way.

The $U(1)$ potentials for general rotating black holes are:
\bea
\beta_H\Phi^j_L &=& {\pi\mu
(\tanh\delta_j 
\prod_i\cosh\delta_i -
\coth\delta_j \prod_i\sinh\delta_i)\over
\sqrt{\mu - (l_1-l_2)^2}}~,
\label{eq:U1L}\\
\beta_H\Phi^j_R &=& {\pi\mu
(\tanh\delta_j \prod_i\cosh\delta_i+\coth\delta_j \prod_i\sinh\delta_i)\over
\sqrt{\mu - (l_1+l_2)^2}}~.
\label{eq:U1R}
\eea
The potentials are important for the description of emission 
processes involving charged 
particles~\cite{fl97,klebanov96,hawking97,dowker}. 
As in the case of rotational velocities we note that it is the
combinations $\beta_H\Phi^{j}_{R,L}$ that can be attributed a given
sector, rather than $\beta_H$ and $\Phi^{j}_{R,L}$ individually.

Finally, from independent scaling symmetries in the two sectors we have the 
sum rules:
\bea
{1\over 2}\beta_R M-\sum_j \beta_H\Phi^j_R Q_j- 
{3\over 2} \beta_H \Omega^R J_R &=& {3\over 2}S_R~, \\
{1\over 2}\beta_L M-\sum_j \beta_H\Phi^j_L Q_j- 
{3\over 2} \beta_H \Omega^L J_L &=& {3\over 2}S_L~,
\eea
that serve as useful checks on the algebra.

\subsection{Spacetime Geometry}
In the preceding subsections the thermodynamic variables were derived
from the entropy; but the standard 
thermodynamic quantities also have direct spacetime interpretations.
The black hole entropy is given in terms of the area of the outer event 
horizon by the Bekenstein-Hawking formula:
\be
S = {A\over 4G_N}~,
\ee 
the physical inverse temperature is defined from the surface 
acceleration $\kappa_{+}$ at the outer event horizon as:
\be
\beta_H= {2\pi\over\kappa_{+}}~,
\ee
and the physical angular velocities are:
\bea
\Omega^R &=& {1\over 2}({d(\phi+\psi)\over dt})_{\rm outer~horizon}~, \\
\Omega^L &=& {1\over 2}({d(\phi-\psi)\over dt})_{\rm outer~horizon}~.
\eea
Direct calculations from the metric indeed verifies that these 
geometric definitions agree with thermodynamics. This 
will be shown in the subsequent section, as a by-product 
of a more detailed exploration.

It is remarkable
that the natural division of thermodynamic 
potentials into independent L and R contributions also 
allows an interpretation in terms of spacetime geometry: 
this follows from the presence of both outer and inner event horizons! 
Indeed, from the area $A_{-}$ of the inner horizon we can define
an ``entropy''\footnote{Variables with index ``$-$'' always denote quantities 
measured at the inner horizon. The corresponding quantities at 
the outer horizon will sometimes be denoted with an index ``+'' 
and sometimes without an index.}:
\bea
S_{-}&\equiv& {A_{-}\over 4G_N} \\
&=&
2\pi [\sqrt{{1\over 4}\mu^3 
(\prod_i \cosh\delta_i+\prod_i \sinh\delta_i)^2-J^2_L}
-\sqrt{{1\over 4}\mu^3 
(\prod_i \cosh\delta_i-\prod_i \sinh\delta_i)^2-J^2_R}]~.
\nonumber
\eea
It follows that~\cite{fl97}:
\be
S_{R,L}={1\over 2}({A_{+}\over 4G_N}\mp {A_{-}\over 4G_N})~.
\ee
Similarly:
\be
\beta_{R,L} = {2\pi\over\kappa_{+}}\pm {2\pi\over\kappa_{-}}~.
\label{eq:betarldef}
\ee
where $\kappa_{\pm}$ are the surface accelerations at the inner and 
outer event horizons, respectively:
\be
{1\over\kappa_{\pm}} =
{{1\over 4}\mu^2 (\prod_i \cosh^2\delta_i-\prod_i \sinh^2\delta_i)
\over
\sqrt{{1\over 4}\mu^3 (\prod_i \cosh\delta_i-\prod_i \sinh\delta_i)^2-J^2_R}}
\pm
{{1\over 4}\mu^2 (\prod_i \cosh^2\delta_i-\prod_i \sinh^2\delta_i)
\over\sqrt{{1\over 4}\mu^3 
(\prod_i \cosh\delta_i+\prod_i \sinh\delta_i)^2-J^2_L}}~.
\label{eq:kappapm} 
\ee
It is suggestive that the 
spacetime geometry divides the entropy and the temperature in 
the {\it very} same way that the microscopic interpretation does.

Next we consider the angular velocities. They are usually
defined from the geometry in the vicinity of the outer event 
horizon. Complementary rotational velocities can be introduced at the 
inner horizon through:
\bea
\Omega^R_{-} &=& {1\over 2}({d(\phi+\psi)\over dt})_{\rm inner~horizon} 
\label{eq:rotvelR}~,\\
\Omega^L_{-} &=& {1\over 2}({d(\phi-\psi)\over dt})_{\rm inner~horizon} 
\label{eq:rotvelL}~.
\eea
However, we have already defined angular momenta $J_{R,L}$ that
couple only to their designated sectors; so in this case it should
not be expected that the rotational velocities would be further 
divided into two contributions. Indeed, in the next section we show that
${1\over\kappa_{-}}\Omega^R_{-}={1\over\kappa_{+}}\Omega^R$ 
and ${1\over\kappa_{-}}\Omega^L_{-}=-{1\over\kappa_{+}}\Omega^L$; 
so the rotational velocities at the inner horizon are not independent 
thermodynamic parameters. (Similar comments apply to the $U(1)$ 
potentials.)

In sum, we find that each thermodynamic variable is split into 
two parts. This is in accord with the microscopic interpretation 
because the string supports both left and right moving excitations; 
and macroscopically it follows as a consequence of the two horizons. 
Note that some special cases have only one event horizon\footnote{These 
include the neutral black holes where one or more of the boost parameters
vanish. An important case is the Schwarzschild black hole.}. However, 
we can interpret these cases as limits that appear when the inner 
horizon coalesces with the curvature singularity, and hence continue
referring to an inner horizon.

\section{General Wave Equation}
\label{sec:waveeq}
A good way to explore the geometry of a black hole is to consider small 
perturbations of the background. The simplest possibility is a 
minimally coupled scalar,
{\it i.e.} a scalar field that satisfies the Klein-Gordon equation:
\be
{1\over\sqrt{-g}}\partial_\mu (\sqrt{-g}g^{\mu\nu}\partial_\nu \Phi) =0~.
\ee
 From the black hole background given in appendix~\ref{app:bhsolution}
it is straightforward to write out the equation explicitly. To present 
the result in a satisfying symmetric form we use the Killing symmetries 
deriving from stationarity, and the two axial symmetries of the rotation
group in four spatial dimensions. Then the wave function can be written:
\be
\Phi\equiv\Phi_0(r)~\chi(\theta)~
e^{-i\omega t+im_\phi\phi+im_\psi\psi}
=\Phi_0(r)~\chi(\theta)~e^{-i\omega t+im_R(\phi+\psi)+im_L(\phi-\psi)}~.
\ee
The angular variables $\phi$ and $\psi$ have period $2\pi$;
so $m_{\phi,\psi}=m_R\pm m_L$ are integer valued.
We also introduce a dimensionless radial coordinate $x$ that is related 
to the standard radial coordinate $r$ through\footnote{More precisely the 
coordinate $r$ 
is the five dimensional
analogue of the Boyer-Lindquist coordinate. It reduces to the 
Schwarzschild coordinate when charges and angular momenta vanish.} :
\be
x \equiv {r^2 - {1\over 2}(r^2_{+}+r^2_{-})\over
(r^2_{+}-r^2_{-})}~.
\label{eq:xdef}
\ee
In this coordinate system the outer and inner horizons at $r_{\pm}$ 
are at $x={1\over 2}$ and $x=-{1\over 2}$, respectively; and the 
asymptotically flat 
region is at $x=\infty$. With this notation the wave equation can 
be written as:
\bea
&~&{\partial\over\partial x}(x^2-{1\over 4}){\partial\over\partial x}\Phi_0
+{1\over 4}[x\Delta\omega^2-\Lambda+M\omega^2+
\label{eq:geneq}
\\
&+&{1\over x-{1\over 2}}
({\omega\over\kappa_{+}}-m_R {\Omega^R\over\kappa_{+}}
-m_L {\Omega^L\over\kappa_{+}})^2
-{1\over x+{1\over 2}}({\omega\over\kappa_{-}}-m_R {\Omega^R\over\kappa_{+}}
+m_L {\Omega^L\over\kappa_{+}})^2]\Phi_0 = 0~. \nonumber
\eea
Here $\kappa_{\pm}$ is the surface acceleration at the inner and
outer event horizon, $\Omega^{R,L}$ are the angular velocities conjugate 
to the two angular momenta, $M$ is the mass, $\Lambda$ is the 
eigenvalue of the angular Laplacian, and $\Delta$ can be expressed 
in terms of the entropy and the temperature as $\Delta=\beta_H^{-1} S$. 
The expressions for $\kappa_{\pm}$ and $\Omega^{R,L}$ are precisely
those given in the preceding section 
(eqs.~\ref{eq:kappapm} and~\ref{eq:omegal}-\ref{eq:omegar}). 
We emphasize that this expression is the exact Klein-Gordon equation 
in the most general black hole background in five dimensions. 
Interestingly it is in fact no more complicated than special cases 
that have been considered previously~\cite{mathur97,strominger97a}.

The wave equation is much simpler than the metric it derives from,
but it nevertheless remains rather involved. Fortunately 
each term has a simple interpretation, as follows:

\paragraph{Energy at infinity:} 
The symbol $\Delta$ can be defined in the equivalent forms:
\be
\Delta \equiv \beta^{-1}_H S = r^2_{+}-r^2_{-}
\ee
When we use the latter form for $\Delta$ and the definition 
of $x$ in terms of the radial variable $r$ (eq.~\ref{eq:xdef}), 
the term ${1\over 4}x\Delta \omega^2$ and the derivative term
in eq.~\ref{eq:geneq} (without the ${1\over 4}$) can be written as:
\be
({1\over r^3}{\partial\over\partial r}r^3 {\partial\over\partial r}
+\omega^2)\Phi_0 = 0~.
\ee
This is simply the radial part of the Klein-Gordon equation in 
five {\it flat} spacetime dimensions. Evidently the term 
${1\over 4}x\Delta \omega^2$ encodes properties of 
the perturbation that persist even in the absence of a black hole.
It can be interpreted physically as the energy of the 
perturbation at infinity. 

We can use the angular momentum parameters $l_{1,2}$
(defined in eq.~\ref{eq:l12def}) to write $\Delta$ as:
\be
\Delta = \sqrt{[\mu-(l_1-l_2)^2][\mu-(l_1+l_2)^2]}~. 
\ee
It is curious that, in terms of  $l_{1,2}$, $\Delta$ does not depend 
on the boost parameters $\delta_i$. 
Note also that this relation shows that, in the absence of angular momentum,
we have simply $\Delta=\mu$ .

\paragraph{The screening terms:}

The term $\Lambda$ reflects the angular momentum barrier. At large
distances it is suppressed relative to the energy at infinity by one 
power of $x\propto r^2$ as expected.
The mass term 
$M$ is the long range gravitational interaction. Coulomb type 
potentials are of the $r^{-2}\propto x^{-1}$ form in five dimensions; 
so it is reasonable that the gravitational screening and the angular momentum 
barrier are of the same order.

The precise form of the angular Laplacian is:
\be
\hat{\Lambda}=
4\vec{K}^2+(l^2_1+l_2^2)\omega^2+(l^2_2-l_1^2)\omega^2\cos 2\theta~,
\label{eq:lambdadef}
\ee
where:
\be
\vec{K}^2 =
- {1\over 4\sin 2\theta}{\partial\over\partial\theta}
\sin 2\theta{\partial\over\partial\theta}
-{1\over 4\sin^2 \theta}{\partial^2\over\partial\phi^2}
-{1\over 4\cos^2 \theta}{\partial^2\over\partial\psi^2}~.
\label{eq:flatlap}
\ee
is the angular Laplacian in five {\it flat} spacetime dimensions.
The rotation of the background modifies the angular momentum barrier
experienced by a small perturbation; but the change is a very mild 
one. Specifically it is $r$ independent so that separation of $\theta$ 
and $r$ variables is still possible. Moreover, it is charge-independent 
when the angular momenta are expressed in terms of $l_{1,2}$. 

\paragraph{The outer event horizon:} Consider the vicinity of 
the outer event horizon $x\sim {1\over 2}$, ignoring temporarily
the angular velocities. On general grounds the geometry of the black 
hole must reduce to Rindler space:
\be
ds^2 = - \kappa^2_{+}\rho^2 dt^2 + d\rho^2 ~.
\label{eq:outermetric} 
\ee
Here $\kappa_{+}$ is clearly identified as the surface acceleration.
The proper radial coordinate $\rho$ is related to the variable $x$ as 
$\rho\sim \sqrt{x-{1\over 2}}$ for $x\sim {1\over 2}$ (with $x>{1\over 2}$).
The solution to the radial wave equation in this regime is of the form:
\be
\Phi_0 \sim e^{-i\omega(t \pm \kappa^{-1}_{+}\log\rho )}
\sim e^{-i\omega(t \pm {1\over 2}\kappa^{-1}_{+}\log (x-{1\over 2}))}~.
\ee
The full wave equation eq.~\ref{eq:geneq} indeed supports solutions of 
this limiting form close to the outer horizon. 
In this way the Rindler space approximation explains the form of 
the singularity at $x={1\over 2}$ in eq.~\ref{eq:geneq}. Specifically
it verifies that the $\kappa_{+}$ of eq.~\ref{eq:geneq} is indeed 
precisely the surface acceleration. Angular parameters can be restored
by transforming to the comoving frame, using the definitions of rotational 
velocities (eqs.~\ref{eq:rotvelR}-\ref{eq:rotvelL}). 
Then the full wave function in this regime becomes:
\be
\Phi\sim e^{-i\omega t+im_R(\phi+\psi)+im_L(\phi-\psi)}
e^{\mp {i\over 2}({\omega\over\kappa_{+}}-m_R {\Omega^R\over\kappa_{+}}
-m_L {\Omega^L\over\kappa_{+}})\log (x-{1\over 2})}\chi(\theta)~.
\ee
Comparison with eq.~\ref{eq:geneq} shows that the rotational parameters 
$\Omega^{R,L}$ have been identified correctly. This constitutes the 
promised verification that the geometrical definition 
of the physical parameters agrees with the thermodynamical one.

For later reference we note that the modes of the form:
\be
\Phi_0^{\rm in} \sim (x-{1\over 2})^{- {i\over 2}
({\omega\over\kappa_{+}}-
m_R {\Omega^R\over\kappa_{+}}
-m_L {\Omega^L\over\kappa_{+}})}~,
\ee
are the infalling modes and those of the form:
\be
\Phi_0^{\rm out} \sim (x-{1\over 2})^{{i\over 2}({\omega\over\kappa_{+}}-
m_R {\Omega^R\over\kappa_{+}}
-m_L {\Omega^L\over\kappa_{+}})}~,
\ee
are outgoing. In general relativity these modes are sometimes referred to
as left and right moving modes, respectively, as this is their direction
in the Rindler diagram. We do not use this terminology here in order to 
avoid confusion with excitations of the effective string.

\paragraph{The inner event horizon:}
Similarly, in the vicinity of the inner event horizon the metric can 
be written:
\be
ds^2 = \kappa^2_{-}\rho^2 dt^2 - d\rho^2~.
\label{eqn:innermetric} 
\ee
Here $\rho$ and $x$ are related as $\rho\sim \sqrt{x+{1\over 2}}$
for $x\sim -{1\over 2}$ (with $x>-{1\over 2}$). 
Note that the overall signature is opposite of the one close to the 
outer horizon (Eq.~\ref{eq:outermetric}). However, the wave equation 
is of second order; so it is unaffected by this change. The modes are: 
\be
\Phi_0 \sim e^{-i\omega(t \pm \kappa_{-}^{-1}\log\rho)}
\sim e^{-i\omega (t \pm {1\over 2}\kappa_{-}^{-1}\log (x+{1\over 2}))}~.
\ee
As before the full wave equation indeed supports modes with this
limiting form close to the inner horizon. Hence, from the approximate
metric close to the inner horizon we understand the form
of the pole term in eq.~\ref{eq:geneq} at $x=-{1\over 2}$,
and verify the physical meaning of the various symbols.
This calculation therefore substantiates the advertized relations between 
thermodynamics and the geometry in the vicinity of the inner horizon.
In particular the relations 
${1\over\kappa_{-}}\Omega^R_{-}={1\over\kappa_{+}}\Omega^R$
and  ${1\over\kappa_{-}}\Omega^L_{-}=-{1\over\kappa_{+}}\Omega^L$
can be read off directly from the inner horizon term.
This explains why the parameter $\kappa_{+}$, associated with
the outer horizon, appears in the pole of the inner horizon: it 
is a consequence of the fact that the $\Omega^{R,L}$ refer 
to quantities at the outer horizon.

\section{Spacetime Symmetries and String Theory}
\label{sec:symmetry}
As we have seen the black hole thermodynamics can be naturally organized 
into an $R$ and an $L$ sector that is related to the black hole event 
horizons; but it is not obvious why they are, roughly, the sum and 
the difference of inner and 
outer horizon contributions. In this section we indicate how this comes 
about, by exhibiting a symmetry of the spacetime geometry that singles 
out precisely these combinations. 

The thermal behavior at the outer horizon can be thought of as a 
complex periodicity of the (real) Rindler time $\tau$. In analogy, 
we introduce a new Rindler-type variable $\sigma$ that encodes the 
complex periodicity close to the inner horizon. Just as 
the ``temperature'' ${\kappa_{-}\over 2\pi}$ of the inner horizon is 
not quite a temperature, because the signature is flipped, the 
variable $\sigma$ is not quite a Rindler ``time'', but rather an analogous 
spatial variable. Introducing these auxiliary variables $\tau$ and $\sigma$ 
directly in the wave equation, and ignoring for the time being
the energy at infinity, the radial part becomes the eigenvalue 
problem:
\be
{\cal H}_r\Phi_0 ={1\over 4}(M\omega^2-\Lambda)\Phi_0~,
\ee
where:
\be
{\cal H}_r = 
-{1\over 4\sinh 2\rho}
{\partial\over\partial\rho}\sinh 2\rho{\partial\over\partial\rho}
-{1\over 4\sinh^2\rho}{\partial^2\over\partial\tau^2}
+{1\over 4\cosh^2\rho}{\partial^2\over\partial\sigma^2}
\label{eq:hrdef}
\ee
is written in terms of the radial variable $\rho$ defined by 
$x={1\over 2}\cosh 2\rho$.
($\rho$ reduces to the proper radial coordinate close to the horizons). 

This radial equation is closely related to an underlying 
$SL(2,R)_R\times SL(2,R)_L$ symmetry group. The generators $\vec{R}$ of 
the $SL(2,R)_R$ group are:
\bea
R_1 &=& {1\over 2}\sin(\tau+\sigma){\partial\over\partial\rho}
+ {1\over 2}\cos(\tau+\sigma)
(\coth\rho
{\partial\over\partial\tau}+\tanh\rho{\partial\over\partial\sigma})~, \\
R_2 &=&-{1\over 2}\cos(\tau+\sigma){\partial\over\partial\rho}
+ {1\over 2}\sin(\tau+\sigma)
(\coth\rho{\partial\over\partial\tau}+
\tanh\rho{\partial\over\partial\sigma})~, \\
R_3 &=& {1\over 2}({\partial\over\partial\tau}+{\partial\over\partial\sigma})
~,
\eea
and the generators $\vec{L}$ of the $SL(2,R)_L$ group are found
by taking $\sigma\rightarrow -\sigma$. The $\vec{R}$ 
satisfy the algebra: 
\be
[R_i, R_j]= i\epsilon_{ijk} (-)^{\delta_{k3}}R_k~,
\ee
and similarly for $\vec{L}$. These are the appropriate commutation relations 
for $SL(2,R)\simeq SO(2,1,R)$. The two sets of generators commute 
$[R_i, L_j]=0$, as they should. It is an important fact that the 
quadratic Casimirs of the groups are identical $\vec{R}^2=\vec{L}^2$ 
and equal to:
\be
\vec{R}^2 = - R^2_1 - R^2_2 + R^2_3 = {\cal H}_r~.
\ee
A maximal set of commuting operators for the $SL(2,R)_R\times SL(2,R)_L$ 
symmetry can be chosen as the two compact generators $R_3$ and $L_3$,
and the quadratic Casimir. The wave function is an eigenfunction of all 
these operators. By abuse of notation we equate the operators and their 
eigenvalues:
\bea
2\pi R_3 &=& \beta_R {\omega\over 2}- \beta_H m_R\Omega^R~, \\
2\pi L_3 &=& \beta_L {\omega\over 2}- \beta_H m_L\Omega^L~, \\
\vec{R}^2 &=&\vec{L}^2={1\over 4}(M\omega^2-\Lambda)~.
\eea
Then the wave function is:
\be
\Phi \sim \Phi_0 e^{R_3(\tau+\sigma)+L_3 (\tau-\sigma)}
\ee
where, as before, $\Phi_0$ denote the radial wave function 
that depends only on $\rho$. The $R_3$ and $L_3$ eigenvalues are the complex
periodicities of the variables $\tau+\sigma$ and $\tau-\sigma$.
They can therefore be thought of as the world sheet temperatures,
if we reinterpret $\tau$ and $\sigma$ as the world sheet variables
of an effective string theory. 

In the calculation just presented we have ignored the term
${1\over 4}x\Delta\omega^2$ of the original wave equation 
(eq.~\ref{eq:geneq}). This term is a property of the perturbing field, 
namely its energy at infinity; so it is possible that the description 
nevertheless indicate the internal structure of the black hole
accurately. The role of the energy at infinity is to ensure that the 
geometry far from the black hole is indeed flat Minkowski space. 
In this sense the troublesome term encodes boundary conditions, and 
so indicates that the internal symmetry $SL(2,R)_R\times SL(2,R)_L$ is 
spontaneously broken. The precise role of the energy at infinity is a 
major concern that must eventually be elucidated.

We conclude this section by exhibiting another symmetry. The exact 
equation (eq.~\ref{eq:geneq}) is invariant under:
\bea
x&\rightarrow &-x~, \\
r^2_{+}& \leftrightarrow & r^2_{-}~~~~~~(\Delta\rightarrow -\Delta )~, \\
2\pi R_3 &\rightarrow & 2\pi R_3~,\\
2\pi L_3 &\rightarrow & -2\pi L_3~.
\eea
Macroscopically this interchanges the role of the two horizons. 
In the microscopic interpretation the symmetry leaves $R_3$ invariant
and acts as a parity transformation on the $L_3$. This is precisely
the way $T$-duality acts on conventional conformal field theories;
so the interchange of horizons can be identified with $T$-duality. From 
this point of view the transformations in spacetime geometry 
generalize the usual $R\rightarrow {\alpha^\prime\over R}$ that 
accompanies $T$-duality in the simplest case.

To avoid misunderstanding we emphasize that the arguments presented in 
this section are entirely in the context of the classical geometry. 
We interpret them as an indication of a strategy towards a comprehensive 
effective string model of black holes, but we do not yet have such a 
model.

\section{Solutions of the Wave Equation}
\label{sec:solutions}

In general eq.~\ref{eq:geneq} is a rather complicated differential 
equation. It has regular singularities at the horizons $x=\pm {1\over 2}$
and an irregular singularity at infinity. The singularity at infinity is 
not of the so-called normal kind; so it can not be cured by absorption
in a determining factor (see eg.~\cite{ince}). The solutions to this kind 
of ordinary differential equation has an essential singularity and it 
is not known how to find them explicitly. However, the equation 
simplifies in various regions of the radial variable $x$.
In the following we consider these cases, postponing the discussion
of their combination into solutions covering all of space
to sec.~\ref{sec:abs}.

We will omit the rotational parameters for simplicity in notation 
but this involves no loss of generality as they can be restored
by the substitutions:
\bea
\beta_R {\omega\over 2} &\rightarrow & 
\beta_R {\omega\over 2}-\beta_H m_R\Omega^R~, 
\label{eq:betarsub} \\
\beta_L {\omega\over 2} &\rightarrow & 
\beta_L {\omega\over 2}-\beta_H m_L\Omega^L~,
\label{eq:betalsub}  \\
\beta_H\omega &\rightarrow & \beta_H\omega-\beta_H m_R\Omega^R- 
\beta_H m_L\Omega_L~.
\eea

\paragraph{The asymptotic region:}
At large $|x|\gg 1$ we approximate eq.~\ref{eq:geneq} by:
\be
{\partial\over\partial x}x^2{\partial\over\partial x}\Phi_0
+{1\over 4}(x\Delta\omega^2-\Lambda+M\omega^2)\Phi_0 = 0
\label{eq:aseq}
\ee
The horizon terms were omitted and we took $x^2-{1\over 4}\simeq x^2$
in the kinetic energy. This equation can be solved exactly 
in terms of Bessel functions. The linearly independent 
solutions are\footnote{For approximate solutions at large distances
we replace the index $0$ of the radial wave functions with $\infty$.}:
\be
\Phi^{\pm}_\infty = 
{1\over x^{1\over 2}}J_{\pm (2\xi-1)} (\omega\sqrt{x\Delta})~,
\ee
where $\xi$ is:
\be
\xi= {1\over 2}(1+\sqrt{1+\Lambda-M\omega^2})~.
\label{eq:xidef}
\ee

\paragraph{The horizon region:}
In the horizon region the wave equation can be approximated by:
\be
[{\partial\over\partial x}(x^2-{1\over 4}){\partial\over\partial x}
\nonumber +{1\over 4}(-\Lambda+M\omega^2+
{1\over x-{1\over 2}}{\omega^2\over\kappa^2_{+}}
-{1\over x+{1\over 2}}{\omega^2\over\kappa^2_{-}})]\Phi_0 = 0~.
\label{eq:horeq}
\ee
The {\it only} approximation is the omission of the term  
${1\over 4}x\Delta\omega^2={1\over 4}x(r^2_{+}-r^2_{-})\omega^2$. 
It is the divergence of this term for large $x$ that is 
responsible for the irregular singularity at infinity in the general 
case; so the approximate equation has three singularities that are 
all regular. This is a
standard problem that is solved by the hypergeometric 
function~\cite{mathur97,strominger97a}. 
One solution is:
\be
\Phi_0^{\rm in}=
({x-{1\over 2}\over x+{1\over 2}})^{-{i\beta_H\omega\over 4\pi}}
(x+{1\over 2})^{-\xi} 
F(\xi-i{\beta_R\omega\over 4\pi},
\xi-i{\beta_L\omega\over 4\pi},
1-i{\beta_H\omega\over 2\pi},{x-{1\over 2}\over x+{1\over 2}})~,
\label{eq:phiin0}
\ee
where $\xi$ was given in eq.~\ref{eq:xidef}. The surface 
accelerations $\kappa_{\pm}$ were eliminated in terms of the 
temperatures $\beta_{R,L}$ and 
$\beta_H={1\over 2}(\beta_R+\beta_L)$ (using eq.~\ref{eq:betarldef}). 
A linearly independent solution can be chosen as:
\be
\Phi_0^{\rm out}=({x-{1\over 2}\over x+{1\over 2}})^{i\beta_H\omega\over 4\pi}
(x+{1\over 2})^{-\xi} 
F(\xi+i{\beta_R \omega\over 4\pi},
\xi+i{\beta_L \omega\over 4\pi},
1+i{\beta_H\omega\over 2\pi},{x-{1\over 2}\over x+{1\over 2}})~.
\ee
The two solutions are related by time reversal. This can be seen
directly by the substitution $\omega\rightarrow -\omega$.

The two independent solutions have been chosen in a form that reflects
the physics in the vicinity of the outer horizon: they reduce to plane 
waves $(x-{1\over 2})^{\pm{i\beta_H\omega\over 4\pi}}$ for 
$x\sim {1\over 2}$. An alternative basis that is adapted to the
behavior at infinity follows by the modular properties of
the hypergeometric functions. For example $\Phi_0^{\rm in}$ of 
eq.~\ref{eq:phiin0} can be written:
\bea
&~&\Phi_0^{\rm in}=
({x-{1\over 2}\over x+{1\over 2}})^{-i\beta_H\omega\over 4\pi}\times \\
&\times & [(x+{1\over 2})^{-\xi}
{\Gamma(1-i{\beta_H\omega\over 2\pi})\Gamma(1-2\xi)\over
\Gamma(1-\xi -i{\beta_L\omega\over 4\pi})
\Gamma(1-\xi -i{\beta_R\omega\over 4\pi})}
F(\xi-i{\beta_R\omega\over 4\pi},
\xi-i{\beta_L\omega\over 4\pi},
2\xi,{1\over x+{1\over 2}}) +\nonumber \\
&+&
(x+{1\over 2})^{\xi-1}
{\Gamma(1-i{\beta_H\omega\over 2\pi})\Gamma(2\xi-1)\over
\Gamma(\xi-i{\beta_L\omega\over 4\pi})
\Gamma(\xi-i{\beta_R\omega\over 4\pi})}
F(1-\xi-i{\beta_L\omega\over 4\pi},
1-\xi-i{\beta_R\omega\over 4\pi},
2-2\xi,{1\over x+{1\over 2}}) ] \nonumber 
\eea
In this form the asymptotic behavior for large $x$ is manifest:
\be
\Phi_0^{\rm in}\sim x^{-\xi}
{\Gamma(1-i{\beta_H\omega\over 2\pi})\Gamma(1-2\xi)\over
\Gamma(1-\xi-i{\beta_L\omega\over 4\pi})
\Gamma(1-\xi-i{\beta_R\omega\over 4\pi})}
+x^{\xi-1}
{\Gamma(1-i{\beta_H\omega\over 2\pi})\Gamma(2\xi-1)\over
\Gamma(\xi-i{\beta_L\omega\over 4\pi})
\Gamma(\xi-i{\beta_R\omega\over 4\pi})}~.
\label{eq:phiininf}
\ee
Here each term admits corrections for large $x$ that are subleading in
${1\over x}$. 

Similarly a basis adapted to the behavior at the inner horizon can be 
chosen. The wave function that has only an ingoing component at the 
outer horizon has both an outgoing and an ingoing component at the 
inner horizon. In physical terms the scattering off the background 
invariably mixes the components. The basis adapted to the inner horizon
will play no role in the present investigation.

\paragraph{The angular Laplacian:}
The angular Laplacian $\vec{K}^2$ of a flat five dimensional background 
(eq.~\ref{eq:flatlap}) is the quadratic Casimir of the group 
$SO(4)\simeq SU(2)_L\times SU(2)_R$. It has eigenvalues 
$\vec{K}^2 = {1\over 4}K(K+2)$ where $K$ is an integer. 
The presence of the curved background modifies the angular Laplacian 
to (eq.~\ref{eq:lambdadef}):
\be
\hat{\Lambda}=
4\vec{K}^2+(l^2_1+l_2^2)\omega^2+(l^2_2-l_1^2)\omega^2\cos 2\theta~.
\ee
The solutions $e^{i(m_\phi \phi + m_\psi \psi)}\chi(\theta)$ to the 
corresponding eigenvalue problem cannot in general be found in closed 
form\footnote{In fact the differential equation is the analytical 
continuation of the radial equation~eq.~\ref{eq:hrdef}:
the constant term is analogous to the mass term and the 
$\cos 2\theta$ term corresponds to the energy at infinity 
(omitted in eq.~\ref{eq:hrdef}).}. As a qualitative result we note that 
the contributions from the rotation of the black hole are always positive. 
In the special case $l_1=l_2$ the eigenfunctions $\chi(\theta)$ are 
hypergeometric functions and the eigenvalues are very simple:
\be
\Lambda = K(K+2) + (l_1^2+l_2^2)\omega^2 
\label{eq:lambdaeigen}
\ee
Corrections can be calculated perturbatively. The leading term is of 
second order in $(l^2_2-l^2_1)\omega^2$ because $\cos 2\theta$ 
vanishes when averaged over all angles. We can use eq.~\ref{eq:lambdaeigen}
as approximate eigenvalues for large classes of problems, including 
those relevant for low energy perturbations, or for black holes 
with nearly coincident rotation parameters.

\section{Absorption Cross-sections}
\label{sec:abs}
The calculation of absorption cross-sections follows much previous
work (including
~\cite{abs1,abs2,greybody,hawking97,dowker,mathur97,strominger97a,emparan}). 
In this section we find the necessary generalizations due to angular 
momentum and sharpen the ranges of validity previously established 
for nonrotating black holes. We first carry out the algebraic manipulations, 
and then consider their ranges of validity.

In the absorption geometry the wave function close to the horizon has
only an incoming component. We normalize the wave function as 
$A_0\Phi_0^{\rm in}$. Then eq.~\ref{eq:phiin0} gives the flux
at the horizon as:
\be
{\rm flux}= {1\over 2i}(\bar{\Phi}\sqrt{g^{rr}}r^3\partial_r\Phi-{\rm c.c.})
=|A_0|^2 {\beta_H\omega\Delta\over 4\pi}~.  
\ee
Similarly, we write the wave function in the asymptotic region as
$A_{\infty}^{+}\Phi_\infty^{+}$ and expand at very large distances:
\be
A_{\infty}^{+}\Phi_\infty^{+} \sim A_\infty^{+} 
\sqrt{2\over\pi x^{3\over 2}\Delta^{1\over 2}\omega}
\cos (\omega\sqrt{x\Delta}-\xi\pi+{1\over 4}\pi)~,
\ee
so the flux becomes:
\be
{\rm flux}= {1\over 2i}(\bar{\Phi} r^3\partial_r\Phi-{\rm c.c.})
= |{A_{\infty}^{+}}|^2 {\Delta\over 4\pi}~.
\label{eq:fluxinf}
\ee
The effective two dimensional transmission coefficient $|T_K|^2$ is 
the ratio of these fluxes. Using a geometric relation derived 
in~\cite{mathur97b} the absorption cross-section of the Kth 
partial wave becomes: 
\be
\sigma^{(K)}_{\rm abs}(\omega) = 
{4\pi (K+1)^2\over\omega^3}|T_K|^2 = 
{4\pi\beta_H\over\omega^2}(K+1)^2 |{A_0\over A_\infty^{+}}|^2~. 
\label{eq:sigmaabs}
\ee
To find the ratio $|A_0/A_\infty^{+}|$ we consider 
a general wave function in the asymptotic region:
\be
\Phi_\infty = 
A_{\infty}^{+}{1\over x^{1\over 2}}J_{2\xi-1} (\omega\sqrt{x\Delta})
+A_{\infty}^{-}{1\over x^{1\over 2}}J_{-{(2\xi-1)}} (\omega\sqrt{x\Delta})~,
\ee
and expand for small arguments of the Bessel function:
\be
\Phi_\infty \sim A_{\infty}^{+} x^{\xi-1}
{1\over\Gamma(2\xi)} ({\sqrt{\Delta}\omega\over 2})^{2\xi-1}
+A_{\infty}^{-} x^{-\xi}
{1\over\Gamma(2-2\xi)} ({\sqrt{\Delta}\omega\over 2})^{1-2\xi}~.
\label{eq:aslim}
\ee
This should be compared with the near-horizon wave function 
$A_0\Phi_0^{\rm in}$ 
for large $x$ (eq.~\ref{eq:phiininf}). Assuming that these limiting 
forms have an overlapping regime of validity we find:
\be
|{A_{\infty}^{+} \over A_0}| =  ({\sqrt{\Delta}\omega\over 2})^{1-2\xi}
       \Gamma(2\xi)\Gamma(2\xi-1){\Gamma(1-i{\beta_H\omega\over 2\pi})\over
\Gamma(\xi-i{\beta_L\omega\over 4\pi})
\Gamma(\xi-i{\beta_R\omega\over 4\pi})}~,
\ee
Note that the ``matching region'' of overlapping validity is necessarily 
at large $x$; so, for $\xi>{1\over 2}$, the $x^{\xi-1}$ terms 
dominate and the $x^{-\xi}$ terms can be neglected. 
This fact was anticipated already in the derivation of the flux 
(eq.\ref{eq:fluxinf}), where $A^{-}_\infty$ was 
ignored\footnote{The case where $\xi$ becomes a complex number corresponds 
to large frequencies. Here both $A_\infty^{-}$
and $A_\infty^{+}$ must be taken into account. In this case the 
appropriate modifications are given in an appendix of~\cite{mathur97}.}. 

Collecting the results the partial absorption cross-sections become: 
\be
\sigma^{(K)}_{\rm abs}(\omega) 
= {4\pi (K+1)^2\beta_H\over\omega^2}
 ({\sqrt{\Delta}\omega\over 2})^{4\xi-2}
|{\Gamma(\xi-i{\beta_L\omega\over 4\pi})
\Gamma(\xi-i{\beta_R\omega\over 4\pi})
\over\Gamma(2\xi-1)\Gamma(2\xi)\Gamma(1-i{\beta_H\omega\over 2\pi})}|^2~.
\label{eq:partabs}
\ee
Recall the definition $\xi= {1\over 2}(1+\sqrt{1+\Lambda-M\omega^2})$
where $\Lambda$ was given in eq.~\ref{eq:lambdadef}.

We turn next to the range of validity for the matching procedure that 
leads to this cross-section. It is most transparent to
derive the conditions directly from the general wave equation 
(eq.~\ref{eq:geneq}) written as:
\bea
&~&{\partial\over\partial x}(x^2-{1\over 4}){\partial\over\partial x}\Phi_0
+
\label{eq:geneqrep} \\
&+&{1\over 4}[x\Delta\omega^2-\Lambda+M\omega^2+
{1\over x-{1\over 2}}({(\beta_R+\beta_L)\over 4\pi})^2 \omega^2
-{1\over x+{1\over 2}}({(\beta_R-\beta_L)\over 4\pi})^2 \omega^2]\Phi_0 = 
0 ~.
\nonumber
\eea
(We assume $m_R=m_L=0$ for convenience, but generality could be restored 
using eqs.~\ref{eq:betarsub}-\ref{eq:betalsub}).
The Bessel function is valid when we can ignore the horizon terms and 
the ${1\over 4}$ in the derivative terms; and the hypergeometric function 
requires that the energy at infinity ${1\over 4}x\Delta \omega^2$ is 
negligible. We must show that there is an intermediate matching 
region where both approximations are valid. We consider two useful
strategies in the following subsections.

\subsection{Matching on a {\it vanishing} potential}
\label{sec:vanish}
The first possibility is that {\it all potential terms are small} in 
the matching region. Then only the kinetic term remains, and the 
equation integrates to a constant solution.
This constant value of the wave function is
the coincident amplitude of the Bessel function at small argument 
and the hypergeometric function at large $x$.
\footnote{The coefficient
of the linearly independent solution, proportional to  $x^{-1}$,
can be determined by matching derivatives. This term contributes 
a flux that is suppressed by $(\Delta\omega^2)^2\ll 1$, due to the 
large matching $x$.}

Matching on a vanishing potential requires a range of $x$ that 
satisfy:
\be
x \gg 1~~;~~~
\Delta x\omega^2 \ll 1~~;~~~
{1\over x}\beta_R\beta_L\omega^2\ll 1~~;~~~
|-\Lambda+M\omega^2|\ll 1~.
\ee
The necessary and sufficient conditions for the existence of such $x$ 
are:
\be
\Delta\omega^2 \ll 1~~;~~~
\beta_R\beta_L \Delta \omega^4\ll 1~~;~~~
|-\Lambda+M\omega^2|\ll 1~.
\label{eq:constcond}
\ee
For higher partial waves a positive integer contributes to 
$\Lambda$ (eq.~\ref{eq:lambdadef}); 
so the last condition can only be satisfied in rather special 
circumstances. In this subsection we only consider the S-wave. 
The last 
condition automatically implies $\xi\simeq 1$; so the coincident
wave function s
in the matching region (eq.~\ref{eq:phiininf} 
or eq.~\ref{eq:aslim}) indeed reduce to constants, as expected.
Moreover, the absorption cross-section takes a particularly simple 
and suggestive form:
\be
\sigma_{\rm abs}^{(0)}(\omega)= 
A |{\Gamma (1-i{\beta_L\omega\over 4\pi})
\Gamma (1-i{\beta_R\omega\over 4\pi})\over 
\Gamma (1-i{\beta_H\omega\over 2\pi})}|^2
=A 
{\beta_L {\omega\over 2}\beta_R {\omega\over 2}\over\beta_H\omega}
{(e^{\beta_H\omega}-1)
\over (e^{\beta_L {\omega\over 2}}-1)(e^{\beta_R {\omega\over 2}}-1)}
\label{eq:vanabs}~,
\ee
where $A$ denotes the area of the black hole. (In rewriting 
eq.~\ref{eq:partabs} we used $\Delta=\beta_H^{-1}S$, $S={1\over 4G_N}A$, 
and $G_N={1\over 4}\pi$.) 
This cross-section can be interpreted microscopically in terms of a 
two-body process of the effective string theory that parametrizes the
collective excitations of the black hole~\cite{mathur,greybody}. 

Note that we have not assumed $\beta_R\omega\sim\beta_L\omega\sim 1$; 
so there are regimes where either one or both of the 
Bose-distribution factors simplify to either the Maxwell distribution
or to the Bose degenerate state. The classical calculation is still 
reliable in  these cases. 

Next we consider some specific examples.

\paragraph{Low energy limit:}
In the S-wave the angular operator $\Lambda\propto\omega^2$; so
for an arbitrary black hole all conditions in eq.~\ref{eq:constcond} 
can be satisfied by taking the energy $\omega$ sufficiently small. 
In this case eq.~\ref{eq:vanabs} applies and the cross-section 
becomes:
\be
\sigma^{(0)}_{\rm abs}(\omega\rightarrow 0)= A~.
\ee
This relation is well-known for scattering off non-rotating black 
holes (see~\cite{gibbons96} and references therein), but the
result here also applies to nonrotating ones.

\paragraph{Two large boosts:}

Assume that two 
of the boost parameters are large, say 
$\delta\equiv\delta_1\sim \delta_2 \gg 1$,
and treat the last one as order unity. 
We generalize this ``dilute gas'' region of Maldacena and 
Strominger~\cite{greybody} by including also large angular momenta with 
$J_R \sim J_L \sim \mu^{3\over 2}e^{2\delta}$ or, 
equivalently, $l_1\sim l_2\sim \mu^{1\over 2}$. ($l_{1,2}$ were defined in 
eq.~\ref{eq:l12def}.) 
In this case $\Delta\sim\mu$, $M\sim\mu e^{2\delta}$, 
$\Lambda\sim\mu\omega^2$, and
$\beta_{R}\sim\beta_{L}\sim\mu^{1\over 2}e^{2\delta}$. 
According to eq.~\ref{eq:constcond} the cross-section eq.~\ref{eq:vanabs}
is reliable 
for frequencies that satisfy $e^{\delta}\mu^{1\over 2}\omega\ll 1$. 
This includes (but is not limited to) the interesting range 
$\omega\sim\beta_{R,L}^{-1}\sim \mu^{-{1\over 2}}e^{-2\delta}$.
The thermodynamic parameters of the absorption cross-section 
eq.~\ref{eq:vanabs} have non-trivial dependence on angular momenta; 
and the inferred distribution functions agree in detail with 
those expected from counting arguments~\cite{rotation1,rotation2}.

\paragraph{Rapidly spinning black holes:}
The freedom provided by the angular momenta also allows for a new kind
of limit: all the boosts are arbitrary but a dilute gas type
region can nevertheless be reached by tuning the angular momentum
parameters so that both inverse temperatures are large. 
This is accomplished by taking $l_2=0$ and tuning 
$\mu-l^2_1=\mu\epsilon^2\ll\mu$ ($l_{1,2}$ were defined in 
eq.~\ref{eq:l12def}.) Then $\Delta\sim\mu\epsilon^2$, 
$M\sim\mu$, $\Lambda\sim\mu\omega^2$, and
$\beta_R\sim\beta_L\sim \mu^{1\over 2}\epsilon^{-1}$. The matching
conditions eq.~\ref{eq:constcond} require $\mu\omega^2\ll 1$. This
range of frequencies includes the interesting ones with 
$\omega\sim\beta_{R,L}^{-1}\sim \mu^{-{1\over 2}}\epsilon$.
Note that in this example {\it no hierarchy in the charges is necessary};
so we capture the entire functional dependence of the temperatures 
on the boost parameters. It is also interesting that in this case 
the black hole is not even approximately supersymmetric.

\paragraph{Near BPS limit:}
We generalize the nonrotating near-BPS black hole (considered 
in~\cite{hawking97,dowker}) by including angular momenta 
$l_1\sim l_2\sim\mu^{1\over 2}$ (This implies 
$J_{R}\sim\mu^{3\over 2}e^\delta$ and $J_{L}\sim\mu^{3\over 2}e^{3\delta}$; 
so there is a hierarchy in the 
angular momenta .) Close to extremality all the boosts are large 
$\delta_i \gg 1$ and we expand systematically 
in $e^{\delta}$ (where $\delta\sim\delta_i$). 
Then $\Delta\sim\mu$, $M\sim\mu e^{2\delta}$,$\Lambda\sim\mu\omega^2$,
$\beta_R\sim\mu^{1\over 2}e^{3\delta}$, 
and $\beta_L\sim\mu^{1\over 2}e^{\delta}$. The conditions 
eq.~\ref{eq:constcond} are satisfied for 
frequencies in the range $\mu^{1\over 2}\omega e^{\delta}\ll 1$.
There is a hierarchy of the temperatures ($\beta_R\gg\beta_L$) in this 
case; so there is no regime where both Bose-factors are significant 
simultaneously. The applicable range of frequencies is
$\omega\sim\beta^{-1}_R\sim\mu^{-{1\over 2}}e^{-3\delta}$ but not 
$\omega\sim\beta^{-1}_L\sim\mu^{-{1\over 2}}e^{-\delta}$; so
only the $\beta_R$ can be reliable probed.

\paragraph{Near-extreme Kerr-Newman limit:}
As the final example we consider the near-extreme Kerr-Newman limit
defined by $\mu-(l_1-l_2)^2 = \mu\epsilon^2 \ll \mu$ (with $l_2\neq 0$). 
Here $\Delta\sim\mu\epsilon$, $M\sim{\Lambda\over\omega^2}\sim\mu$, 
$\beta_L\sim\mu^{1\over 2}$, and  $\beta_R\sim\mu^{1\over 2}\epsilon^{-1}$; 
so the condition on the frequency becomes $\mu\omega^2\ll 1$. As in 
the near BPS case we can probe $\beta_R$, but not $\beta_L$.

It is interesting that in the limit $\epsilon\rightarrow 0$ the entropies 
approach $S_R=0$ and:
\be
S=S_L = 2\pi\sqrt{Q_1 Q_2 Q_3 + J^2_R - J^2_L}
= 2\pi\sqrt{n_1 n_2 n_3 + J^2_R - J^2_L}~.
\ee
where the $n_i$ are quantized charges. The near-extreme Kerr-Newman 
limit is not supersymmetric, but the form of the entropy is 
nevertheless reminiscent of the BPS case: the entropy does not depend
on moduli, and the counting arguments can be made notably less
heuristic. 

\subsection{Matching on a {\it constant} potential:}
In this case {\it the screening term} dominates in the matching region.
Then the wave equation is solved by the polynomials $x^{\xi-1}$ and 
$x^{-\xi}$. The coincident wave functions 
(eq.~\ref{eq:phiininf} or eq.~\ref{eq:aslim}) indeed reduces to precisely
these polynomials.

Matching on a constant potential requires a range of $x$ 
so that:
\be
x \gg 1~~;~~~
x\Delta \omega^2 \ll |-\Lambda+M\omega^2|~~;~~~
{1\over x}\beta_R\beta_L\omega^2\ll |-\Lambda+M\omega^2|~.
\label{eq:altcond}
\ee
If $|-\Lambda+M\omega^2|\ll 1$ the present procedure corresponds to
matching on a vanishing potential; but in this case the conditions 
eq.~\ref{eq:altcond} are nevertheless stronger than eq.~\ref{eq:constcond},
because here we insist that the screening term dominates even though 
it is small when $|-\Lambda+M\omega^2|\ll 1$ . Therefore the two matching 
procedures must be considered separately to find the most generous 
ranges of validity. 

The necessary and sufficient conditions for the existence of $x$ 
satisfying eq.~\ref{eq:altcond} are:
\be
\Delta\omega^2 \ll |-\Lambda+M\omega^2|~~;~~~
\beta_R\beta_L \Delta \omega^4\ll  |-\Lambda+M\omega^2|^2~.
\ee
In the S-wave $\Lambda\propto\omega^2$; so in this case there are 
{\it no} assumptions about the frequency of the radiation\footnote{Note
however that we only give the final result for $\xi>{1\over 2}$; but 
the argument shows that the analogous calculation for 
$\xi$ complex is reliable as well.}. Indeed, in the S-wave the entire
potential in eq.~\ref{eq:geneqrep} is proportional to $\omega^2$;
so conditions on the relative size of potential terms must
be frequency independent.

We consider a few specific examples.

\paragraph{Higher angular momentum modes:}
The simplest example of matching on a constant potential concerns 
a particular partial wave $K$, but otherwise the same 
restrictions as in the case of matching on a vanishing potential. 
This is consistent with eq.~\ref{eq:altcond}
(but not eq.~\ref{eq:constcond}).
In this case $\Lambda\simeq K(K+2)$ and $M\omega^2\ll 1$ so
the absorption spectrum is eq.~\ref{eq:partabs}
with $\xi={K\over 2}+1$.
The process can be 
modelled microscopically as an impinging closed string that is absorbed 
by bound state of D-branes, with $2K$ fermions being excited in the 
process~\cite{strominger97a,mathur97b,gubser}.

\paragraph{One large charge:}
We consider the S-wave and take 
$\delta\equiv\delta_3 \gg 1$ and $\delta_{1,2}$ of order 1~\cite{mathur97}. 
Angular momenta $l_{1,2}\sim\mu^{1\over 2}$ can be included.
Then $\Delta\sim\mu$, $M\sim\mu e^{2\delta}$, $\Lambda\sim\mu\omega^2$,
and $\beta_R\sim\beta_L\sim\mu^{1\over 2}e^{\delta}$. This is sufficient
to satisfy the conditions $\Delta\ll M$ and $\beta_R\beta_L\Delta\ll M^2$ 
required by eq.~\ref{eq:altcond}; so the absorption cross-section is given 
by eq.~\ref{eq:partabs} with a general value of $\xi$.

\section{Discussion}
\label{sec:discussion}
We would like to conclude the paper with remarks on the microscopic 
interpretation of our results. As a starting point for the
discussion we consider the Hawking emission rate:
\be
\Gamma^{(0)}_{\rm em}(\omega)=\sigma_{\rm abs}(\omega)~
{1\over e^{\beta_H\omega}-1}~{d^4 k\over (2\pi)^4}~.
\ee
In the regime where matching on a vanishing potential is justified 
(eq.~\ref{eq:constcond}) we use eq.~\ref{eq:vanabs} for the 
cross-section and find:
\bea
\Gamma^{(0)}_{\rm em}(\omega)&=&A~
{\beta_L {\omega\over 2}\beta_R {\omega\over 2}\over\beta_H\omega}~
{1\over (e^{\beta_L {\omega\over 2}}-1)(e^{\beta_R {\omega\over 2}}-1)}
~{d^4 k\over (2\pi)^4} \\
&=&
8\pi G_N~{\cal L}~{1\over\omega}~({\omega\over 2})^2~
{1\over (e^{\beta_L {\omega\over 2}}-1)(e^{\beta_R {\omega\over 2}}-1)}
~{d^4 k\over (2\pi)^4}~,
\label{eq:gammaem}
\eea
where the intermediate step used relations given in 
sec.~\ref{sec:thermo}; and we defined ${\cal L}$ as:
\be
{\cal L} = 2\pi\mu^2 (\prod_i \cosh^2\delta_i -\prod_i \sinh^2\delta_i)~.
\label{eq:leff}
\ee 
It was shown by Das and Mathur that the emission rate eq.~\ref{eq:gammaem} 
is identical, including the coefficient, to the two-body annihilation rate 
for small amplitude waves propagating on an effective string of length 
${\cal L}$~\cite{mathur}. In this model of the emission process the length of 
the effective string parametrizes the strength of the interactions. 
It is satisfying that in our case the length ${\cal L}$ is both U-duality 
invariant and independent of angular momenta. 

For large black holes ${\cal L}$ is much larger that the naive 
string length. The importance of this kind of ``tension renormalization'' 
was recognized already in the early countings of non-perturbative string 
states~\cite{speculations,structure,cfthair1}; and it is now 
understood from D-brane properties how this may come 
about~\cite{fatbh,dmvv}. The near-BPS black holes related to
momentum carrying bound states of D1- and D5-branes~\cite{callan96a,strom96b} 
are special cases of the general formula eq.~\ref{eq:leff}: here
two boosts 
are large $\delta_1\sim\delta_2\gg 1$ and the length reduces to 
${\cal L}=2\pi Q_1 Q_2=2\pi n_1 n_2 R$, where $n_{1,2}$ are the 
quantized D1- and D5-brane charges and $R$ is the length of the 
dimension that the D1-brane wraps around~\cite{mathur}. However,
the general expression for ${\cal L}$ accounts for emission from a larger 
class of black holes than has previously been considered. For example
the full dependence on boost parameters is needed in the case of 
rapidly spinning black holes even though the thermodynamic properties
of this case are analogous to the ``dilute gas'' regime 
of~\cite{greybody}. 

In the microscopic interpretation the colliding quanta have 
Bose-distributions with inverse temperatures 
(eqs.~\ref{eq:betal}-\ref{eq:betar}):
\be
\beta_{R,L}= 
{2\pi\mu (\prod_i \cosh\delta_i\pm\prod_i \sinh\delta_i)\over
\sqrt{\mu - (l_1\pm l_2)^2}}~.
\ee
The dynamical considerations therefore give direct 
information about properties of the microscopic theory. In 
particular, this gives a concrete physical meaning to the
temperatures derived at each event horizon. However, the 
two-body form of the emission rate is a low energy approximation; 
so only the cases where the precise requirement (eq.~\ref{eq:constcond}) 
on the frequency is consistent with the interesting ranges 
$\omega\sim\beta_R^{-1}$ and $\omega\sim\beta_L^{-1}$ can be probed 
in detail~\cite{greybody}. Despite this restriction we can verify the 
dependence of the inverse temperatures on all boost 
parameters by considering rapidly rotating black holes. Our expressions 
for the $U(1)$ potentials (eqs.~\ref{eq:U1L}-\ref{eq:U1R})
can similarly be checked in some regimes, by considering emission of 
charged particles, and the angular potentials 
(eqs.~\ref{eq:omegal}-\ref{eq:omegar}) can be probed by 
considering the emission of higher partial waves\footnote{This 
calculation uses matching on a constant potential, not a vanishing
one.}. Hence the microscopic model based on the thermodynamics of 
two horizons provides an economical summary of a large class of special 
cases, including some that have not been considered before.

The Hawking emission process can be described as a two-body process
in the entire regime where matching on a constant potential is 
justified (eq.~\ref{eq:constcond}). 
For generic non-extremal black holes this implies 
$\beta_{L,R}\omega\ll 1$; so the agreement between the microscopic 
model and the macroscopic calculation reduces to a single number, 
namely the universal low-energy absorption cross-section. 
This is nevertheless non-trivial because we consider the most general 
black holes and the model captures the full functional dependence on
all parameters. It has previously been argued (along somewhat
different lines) that the universal low energy scattering off
Schwarzschild~\cite{susskind96} and 
Reissner-Nordstr\"{o}m~\cite{halyo96} black holes can be accounted
for by an effective string model. 
Our result includes these observations as special cases as well as the 
D-brane inspired string models for near BPS-black holes. Let us 
summarize the argument: from the horizon structure 
we identify distribution functions for right and left moving string 
excitations, from rapidly spinning black holes we infer the coupling 
between the two sectors; and then a calculation gives the universal 
low-energy cross-section for {\it all} black holes. 
In this sense the version of the effective string model presented in 
this paper has some applicability even for generic non-extremal black 
holes.

The remaining problem becomes one of interactions, rather than 
that of state counting. Here it is concerning that in general the 
typical Hawking particle is too energetic to result from a simple 
two-body process. This may simply indicate that interactions are more 
involved at larger energies, at least in the range of parameters where 
matching on a constant potential is justified 
(eq.~\ref{eq:altcond})~\cite{mathur97}. Here the absorption 
cross-section (eq.~\ref{eq:partabs}) depends on the parameter 
$\xi= {1\over 2}(1+\sqrt{1+\Lambda-M\omega^2})$. The angular momentum 
eigenvalue $\Lambda$ (eq.~\ref{eq:lambdaeigen}) depends on the angular 
momentum of the particle as well as that of the background. When the 
main contribution to $\xi$ is from particle angular momentum 
the $\xi$ is integer or half-integer and the spectrum 
can be understood qualitatively from many-body 
kinematics~\cite{strominger97a,mathur97b,gubser}. In general the 
background mass and angular momenta contribute to $\xi$ but the 
emission spectrum retains its qualitative character. It is therefore 
reasonable to suspect that further understanding 
of many-body effects might account also for this case.

As we saw in sec.~\ref{sec:symmetry} the geometry of the region 
in the vicinity of the horizons immediately suggests an effective 
description in string theory. The matching on a vanishing potential 
corresponds to the situation 
where this suggestive near-horizon region can be unambiguously 
distinguished from the surrounding space. In the case of matching on 
a constant potential the long range fields make the distinction less
clear, but presumably still valid, as we argued in the previous paragraph. 
However, in the most general problem the distinction seems ambiguous; 
and it is the processes that are 
sensitive to this coupling between the near-horizon region and the
asymptotic space that we are presently unable to 
account for even classically\footnote{It is possible that investigations 
involving particles with non-minimal coupling (initiated in~\cite{cgkt})
might help, for example by being less sensitive to the term at infinity.}. 
This seems to be a barrier that will remain difficult to surmount in the 
string theory description. It is not yet clear whether this represents an 
obstacle of purely technical nature, or a more profound crisis.

\vspace{0.2in} {\bf Acknowledgments:} 
We would like to thank V. Balasubramanian, C. Callan, S. Mathur, 
H. Verlinde, A. Peet, and F. Wilczek for discussions; and 
D. Kastor and J. Maldacena for correspondence.
This work is supported in part by DOE grant DOE-EY-76-02-3071, NSF
Career advancement award PHY95-12732 (MC), and NATO collaborative
grant CGR 949870 (MC).

\appendix
\section{The Black Hole Solution}
\label{app:bhsolution}

The Einstein metric of the black holes is~\cite{cy96a}:
\bea
{\bar\Delta}^{-{1\over 3}}ds^2_E &=& 
- {(r^2 + l^2_1\cos^2\theta+ l^2_2\sin^2\theta)
(r^2 + l^2_1\cos^2\theta+ l^2_2\sin^2\theta-2m)\over {\bar\Delta}}dt^2 
\nonumber \\
&+&{r^2\over (r^2+l^2_1)(r^2+l^2_2)-2mr^2}dr^2 + d\theta^2+ 
\nonumber \\
&+& {4m\cos^2\theta\sin^2\theta\over{\bar\Delta}}
[l_1 l_2 \{ (r^2 + l^2_1\cos^2\theta+ l^2_2\sin^2\theta)
-2m\prod_{i<j}\sinh^2\delta_i\sinh^2\delta_j \} 
\nonumber \\
&+&2m (l^2_1+l^2_2)\prod_i \cosh\delta_i\sinh\delta_i
-4ml_1 l_2\prod_i\sinh^2\delta_i ]d\phi d\psi-
\nonumber \\
&-&{4m\sin^2\theta\over{\bar\Delta}}
[(r^2 + l^2_1\cos^2\theta+ l^2_2\sin^2\theta)
(l_1\prod_i\cosh\delta_i-l_2\prod_i\sinh\delta_i) +
\nonumber \\
&+&2ml_2 \prod_i\sinh\delta_i ]d\phi dt- \nonumber \\
&-&{4m\cos^2\theta\over{\bar\Delta}}
[(r^2 + l^2_1\cos^2\theta+ l^2_2\sin^2\theta)
(l_2\prod_i\cosh\delta_i-l_1\prod_i\sinh\delta_i) +\nonumber \\
&+&2ml_1\prod_i\sinh\delta_i ]d\psi dt +\nonumber \\
&+& {\sin^2\theta\over{\bar\Delta}}[
(r^2 + 2m\sinh^2\delta_3 + l^2_1)
(r^2 + 2m\sinh^2\delta_1 + l^2_1\cos^2\theta+l^2_2\sin^2\theta)\times
\nonumber \\
&\times& (r^2 + 2m\sinh^2\delta_2 + l^2_1\cos^2\theta+l^2_2\sin^2\theta) 
+\nonumber \\
&+&2m\sin^2\theta\{ (l^2_1\cosh^2\delta_3 -l^2_2\sinh^2\delta_3)
(r^2 + l^2_1\cos^2\theta+l^2_2\sin^2\theta) +\nonumber \\
&+&4ml_1 l_2 \prod_{i<j}\cosh\delta_i\sinh\delta_j 
-2m\sinh^2\delta_1 \sinh^2\delta_2 (l^2_1\cosh^2\delta_3
+l^2_2\sinh^2\delta_3) -\nonumber \\
&-&2ml^2_2 \sinh^2\delta_3 (\sinh^2\delta_1+\sinh^2\delta_2)\}]d\phi^2+
\nonumber \\
&+& {\cos^2\theta\over{\bar\Delta}}[
(r^2 + 2m\sinh^2\delta_3 + l^2_2)
(r^2 + 2m\sinh^2\delta_1 + l^2_1\cos^2\theta+l^2_2\sin^2\theta)\times
\nonumber \\
&\times& (r^2 + 2m\sinh^2\delta_2 + l^2_1\cos^2\theta+l^2_2\sin^2\theta) 
+\nonumber \\
&+&2m\cos^2\theta\{ (l^2_2\cosh^2\delta_3 -l^2_1\sinh^2\delta_3)
(r^2 + l^2_1\cos^2\theta+l^2_2\sin^2\theta) +\nonumber \\
&+&4ml_1 l_2 \prod_{i<j}\cosh\delta_i\sinh\delta_j 
-2m\sinh^2\delta_1 \sinh^2\delta_2 (l^2_2\cosh^2\delta_3
+l^2_1\sinh^2\delta_3) -\nonumber \\
&-&2ml^2_1 \sinh^2\delta_3 (\sinh^2\delta_1+\sinh^2\delta_2)\}]d\psi^2
\eea
where:
\be
{\bar\Delta} = \prod_i (r^2 + 2m\sinh^2\delta_i + l^2_1\cos^2\theta
+l^2_2\sin^2\theta)
\ee
The notation follows~\cite{cy96a}, except that the indices
on the boosts $\delta$ have been redefined $(e1,e2,e)\rightarrow (1,2,3)$.
The $\mu$ of the main text is related to $m$ through $\mu = 2m$. 
Note that the complete solution also includes gauge fields and
other matter fields (of considerable complexity). They are given 
in~\cite{cy96a}. 

It is possible that the metric can be written in a more compact and 
symmetrical form, but we are not aware of any substantial simplifications. 
One helpful identity (that is non-trivial to verify) is:
\be
\sqrt{-g} = r{\bar\Delta}^{1\over 3}\sin\theta \cos\theta
\ee
We inverted the metric using this relation repeatedly and, after 
lengthy manipulation of the resulting formulae, found 
certain complete squares in the resulting wave equation. These are 
the terms that are recognized as the horizon terms in the general 
equation (eq.~\ref{eq:geneq}), after the linear change of radial variable 
(eq.~\ref{eq:xdef}).

\end{document}